# Ultralow-threshold green fluorescent protein laser based on high Q microbubble resonators

SHUOYING ZHAO,[1,2] GAOSHANG LI,[1,2] XUBIAO PENG,[1,2] JIYANG MA, [1,2*] ZHANGQI YIN, [1,2] AND QING ZHAO [1,2]

[1]*Center for Quantum Technology Research and Key Laboratory of Advanced Optoelectronic Quantum Architecture and Measurements (MOE), School of Physics, Beijing Institute of Technology, Beijing 100081, China*
[2]*Beijing Academy of Quantum Information Sciences, Beijing 100193, China*
*mjy@bit.edu.cn

**Abstract:** Biological lasers have attracted vast attention because of their potential medical application prospects, especially the low threshold biological laser, which can be used for ultrasensitive biological detection while ensuring that its luminous gain medium is not damaged by the high-energy pump light. By coupling the low concentration green fluorescent protein (GFP) solution with a high Q whispering gallery mode microbubble resonator, we managed to fabricate a miniature GFP laser with ultralow lasing threshold of 500 nJ/mm$^2$. The energy used to excite the GFP can be reduced to 380 fJ, two orders of magnitude lower than that of the lowest excitation energy GFP laser known. The Q value of the optical cavity in this biological laser is $5.3 \times 10^7$, the highest among GFP lasers at present. We further confirmed the long-term stability of the working characteristics of GFP laser for the first time and found that its optical characteristics can be maintained for at least 23 days. Finally, we measured the effects of different concentrations of fluorescent protein on the laser threshold. The data show that this biological laser can be used for a highly sensitive detection of GFP concentration.

## 1. Introduction

Biolaser, a novel type of laser has garnered tremendous attention in the last several years [1-10]. It utilizes biomaterials as the optical gain medium with several advantages, such as morphological flexibility, low toxicity, and high biocompatibility [11-13]. Compared with the fluorescence emission of biomaterials, biolaser bears the superior characteristics, including but not limited to narrow linewidth, high signal-to-noise ratio, and threshold behavior [1,2]. These features can find applications in biosensing [14-20], bioimaging [21], and so on.

Fluorescent protein [22], an ideal light emitting material covering the entire visible light band, has gained considerable interest in the last few decades, thus playing a significant role in several fields, such as biomedicine and biosensing. Among the fluorescent proteins, green fluorescent protein (GFP) [23], found in Victoria jelly fish in 1962, was studied for the first time. It has high quantum yields of about 0.8 and large transition cross section of $2 \times 10^{-16}$ cm$^2$ [24]. Besides, GFP is easily produced and genetically programmed in nearly all organisms. GFP has a unique molecular structure with the fluorophore enclosed by a β barrel separating the light emitting center by a sufficient distance which can prevent self-quenching during fluorescence, making it one of the most compelling and active candidates in biological research. GFP biolasers have been developed for over ten years [24-31] with different types of resonators, including Fabry–Perot [24-26], whispering gallery mode [27-29], distributed feedback [30,31], etc. GFP biolaser can even be generated in a single cell [24]. The form of GFP in these lasers also differs significantly, including liquid solution [24, 29], single molecule layer [28] and solid state

[25-27, 30, 31]. However, all such lasers suffer from various kinds of problems. Low Q factor (typically below $10^5$) of the resonators [24-31] lead to broad laser linewidth and high threshold pump energy [24-26, 28-31]. While some solid state GFP lasers with extremely high GFP concentrations (>>100 μM) possess low threshold energy [27], it is not friendly to the liquid environment in nearly all the organisms. Also, most of the lasers bear large size [24-27, 30, 31]. These drawbacks, to a large extent, limit its potential applications in biosensing and biomedical analysis.

In addition, preserving the luminescence gain medium for a long time is difficult for the current fluorescent protein laser because of the high threshold and resonator shape limitation. According to current literature, no fluorescent protein laser can be well preserved for more than one day [6, 24-31]. This fatal defect restricts the practical application prospect of biological laser in the future.

In this paper, we couple low concentration GFP solution (typically 10 μM) with a high Q microbubble resonator and generate low threshold biolaser. Microbubble resonator [32], a whispering gallery mode cavity [33, 34], is a miniature hollow sphere made from the silica capillary with thin-walled periphery sustaining the WGM optical field. The hollow structure forms a microfluidic tunnel, facilitating the injection of different optical fluorescent materials [35]. The distinct evanescent optical field inside the microbubble facilitates the interaction between the microresonator and the optical gain medium. The microbubble used in our work has an diameter about 100 μm and exhibits the Q factor of $5.3 \times 10^7$, the highest Q factor of all the GFP laser to date, as per our knowledge. Due to the high Q factor, the GFP laser threshold can reach as low as 500 nJ/mm$^2$. Considering the ultralow mode volume of the microbubble resonator, the pump energy required to excite the GFP interacting with the optical field of the resonator is only 380 fJ, two orders lower than the threshold of existing lasers [24-31]. Furthermore, we test the high durability of the GFP laser, which can last for at least 23 days, and develop a GFP concentration detector based on the threshold behavior of the laser.

## 2. Experiment Setup and Sample Preparation Method

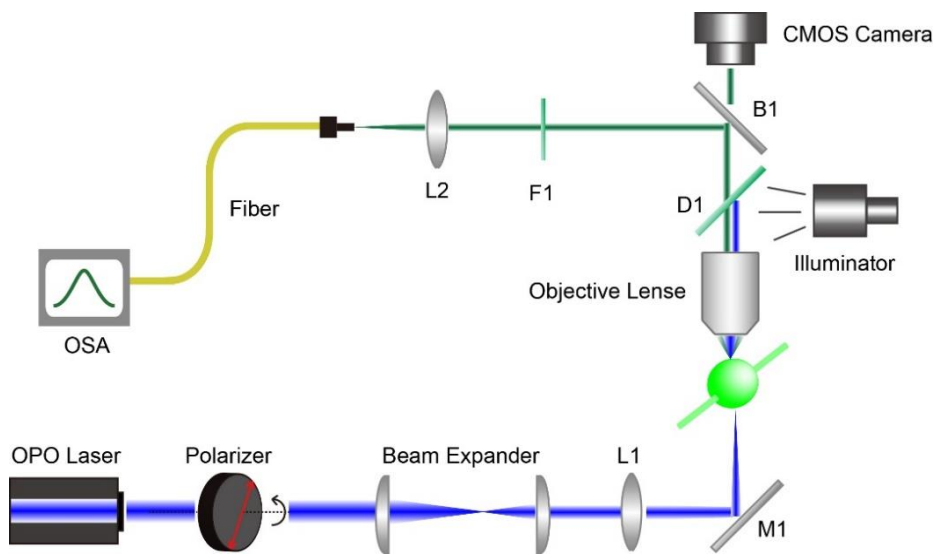

Fig. 1. Experiment setup. Acronyms in the figure: L1, L2: convex lens; M1: flat mirror; B1: 50/50 beam splitter; D1: dichroic lens that can transmit green light and reflect blue light; F1: long-pass filter that can transmit green light and block blue light; OSA: optical spectrum analyzer

The experiment setup is illustrated in Fig. 1. An OPO laser (7 ns pulse duration, 10 Hz repetition rate) is used as the pump laser and the excitation wavelength for GFP is 473 nm. Its energy is controlled by a polarizer, and the beam diameter is reduced by a beam expander. After transmitting through a convex lens L1 (20 cm focal length) and a flat mirror, the OPO laser radiates at the microbubble. The spot size diameter of OPO laser on the microbubble is 1 mm, large enough to cover the whole microbubble. The GFP laser from the microbubble is collected by a 20× objective lens with a N.A. of 0.4. The dichroic mirror D1 is used to transmit GFP laser (around 520 nm) and filter out the OPO laser (473 nm). After passing through a 50/50 beam splitter B1 and an optical long-pass filter F1 to further filter out the pump laser, the GFP laser is focused by a convex lens L2 and collected by a multimode fiber, which guides the laser to the optical spectrum analyzer (Ocean Optics Maya2000 Pro, 1-nm resolution). The illuminator and the CMOS camera monitor the microbubble position, ensuring its radiation by the pump laser.

The microbubble fabrication process can be described as follows: First, the silica capillary (Poymicro Technologies TSP100170, outer diameter of 140 μm, inner diameter of 100 μm, wall thickness 20 μm) is cleaned up by piranha solution at 155°C. Second, the capillary is etched with 5% hydrofluoric acid to minimize the outer diameter of the capillary to 122 μm with the corresponding wall thickness of 11 μm. Then, the capillary is drawn from both sides while heating up with the hydrogen flame to become thinner until its outer diameter becomes approximately 50 μm with the wall thickness of 4.5 μm. Next, one end of the capillary is sealed with the ultraviolet (UV) glue, and pressure is added into the capillary from the other end using a syringe pump. At the same time, the thinnest part of the capillary is irradiated by a $CO_2$ laser. Due to the high absorption of $CO_2$ laser by silica, the capillary softens and blows up to form a microbubble [36].

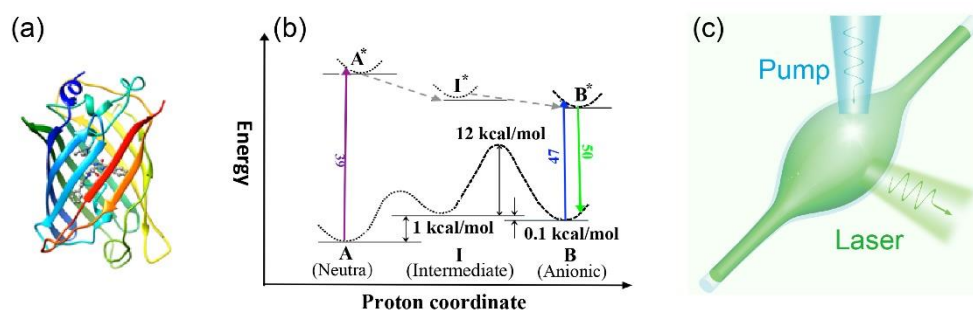

Fig. 2. (a) 3D helical structure of green fluorescent protein (GFP); (b) Energy level structure of GFP. Reprinted from [37]. (c) Schematic diagram of GFP laser

The GFP is prepared according to the following standard scheme: the PET28a-EGFP expression plasmid is transformed into E. coli BL21 (DE3) and cultured on Luria Bertani (LB)-agar plate with antibiotic (kanamycin: 50 μg/ml) at 37°C for 14 h. The positive clone strains are inoculated in the LB liquid medium with kanamycin (50 μg/ml), cultured on a 220 r/min shaking table at 37°C until A280 nm = 0.4, induced by isopropyl thiogalactoside (IPTG), and the colony fluorescence is evaluated after culturing at 25°C for 12 h. Then, the bacteria are cleaved by ultrasound. The GFP protein is purified by nickel resin column under non denaturation conditions. The expressed protein is identified by 12 sodium sulfate polyacrylamide gel electrophoresis (SDS-PAGE). Figs. 2(a) and (b) [37] are the 3D spiral structure and energy level structure diagrams of GFP, respectively. It can be seen from the figure that the multilevel structure and isomerization of GFP provide favorable conditions for light amplification, currently one of the biggest advantages of using organic matter as laser gain medium.

## 3. Results

In the first experiment, we demonstrate an ultralow threshold GFP laser as schematically shown in Fig. 2(c). The microbubble in the experiment has an outer diameter of 110 μm and wall thickness of 1.5 μm [Fig. 3(a)]. The fabrication method of the microbubble is described in the last section. The concentration of the GFP solution is 10 μM, and we inject it into the microbubble by a syringe pump. According to the simulation result [Fig. 3 (b)], the optical mode of the microbubble has a distinct evanescent field inside the hollow core suitable for the interaction between the resonator and the GFP solution. The Q value is measured by coupling the microbubble with a tapered single mode fiber (~1 μm diameter) and scanning the wavelength of the tunable laser (New Focus TLB 6712-P) through the fiber. We can determine the Q factor by testing the full width at half maximum of the Lorentz shaped transmission line. This method has been specifically described in the former studies [38, 39]. The initial Q factor is $10^8$ at 775 nm, and can still retain a high level at $5.3 \times 10^7$ after filling with GFP solution [Fig. 3(c)]. It has the highest Q factor than any other GFP laser. Fig. 3(d) shows the mode distribution of the microbubble in an FSR separation.

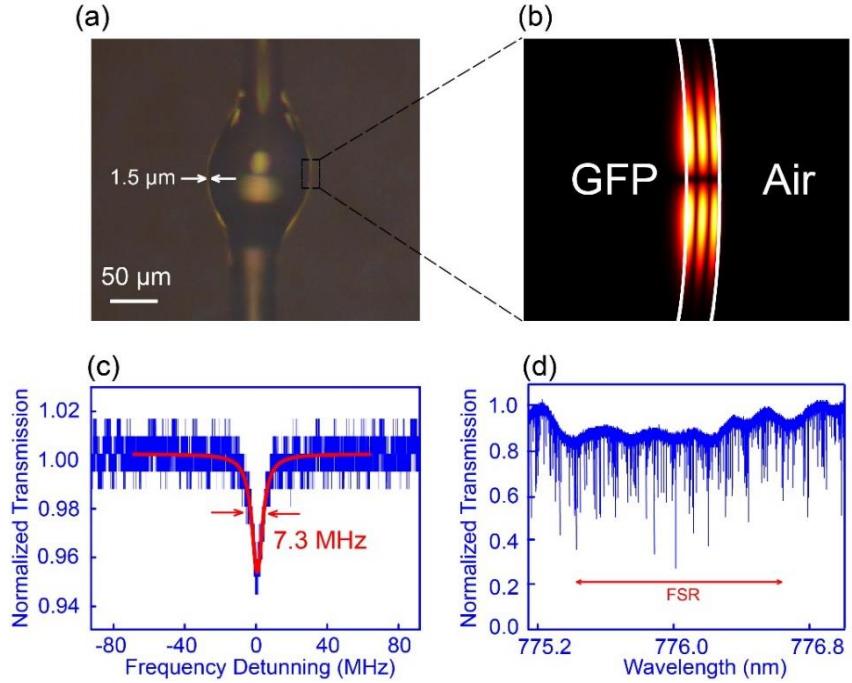

Fig. 3. (a) Microscopic image of the microbubble; (b) Simulation results of the optical mode of the microbubble; (c) Experimental measurement of the Q factor of the microbubble at 775 nm, the linewidth is 7.3 MHz, corresponding to the Q factor of $5.3 \times 10^7$; (d) Mode distribution of the microbubble in an FSR separation

We use the setup described in the last section to generate and collect the GFP laser. Fig. 4 (a) shows the optical spectrum at different pumping energies; according to the Q factor of the microbubble, the linewidth of laser should be at the level of $1.1 \times 10^{-5}$ nm, the seemingly broad linewidth is attributed to the multimode feature of the microbubble resonator and the limited resolution (~1 nm) of the optical spectrum analyzer. As specified in Fig. 3 (d), wavelength separation between the neighboring modes is approximately 0.01 nm, far beyond the resolution of the optical spectrum analyzer. The laser threshold measurement is performed by testing the peak energy of the optical spectrum at different pump energies, as illustrated in Fig. 4 (b). A kink appears on plotting the peak energy with the variation in the pump energy, an evident proof of laser generation with the threshold of 500 nJ/mm$^2$. Only the GFP in the evanescent field (see Fig.3 (b)) in the hollow area of the microbubble can strongly interact with the microbubble and contribute to lasing. According to simulation result, the mode volume in this area is $5.62 \times 10^{-17}$ m$^3$, which is 8/100000 of the whole volume of the microbubble, the threshold pump energy on the whole microbubble is 4.75 nJ, so 380 fJ energy is used to excite the GFP in the evanescent field of the microbubble resonator. To the best of our knowledge, this is the lowest threshold in any known type of GFP laser. This low threshold characteristic indicates that we can excite sufficiently strong optical signals for detection with low energy pump light to avoid the damage of organisms and luminescent materials via high pump light irradiation, which is of great significance to expand the service life of GFP laser and its application in the biological detection.

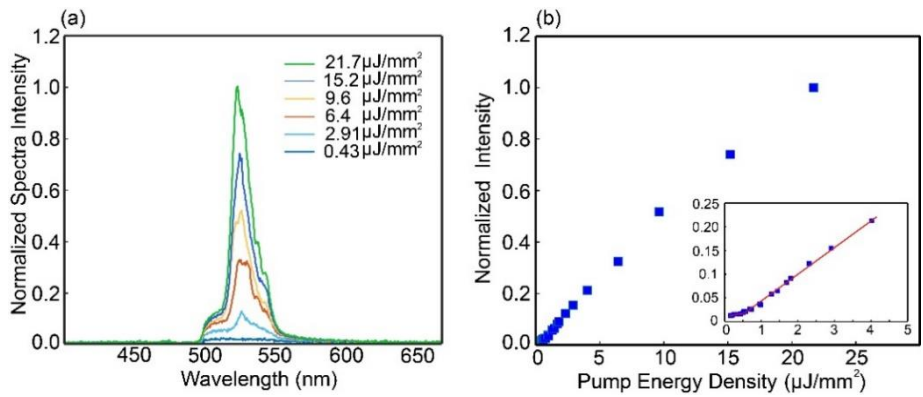

Fig. 4. (a) Lasing spectra under different pump energy densities; (b) Normalized laser intensity with the variation of pump energy density, inset: close-up of the low pump energy regime with the threshold of 500 nJ/mm$^2$

Then, we test the durability of the GFP laser by investigating its threshold behaviors in the separation lasting for several days. In this experiment, the microbubble resonator with diameter and wall thickness of 115 μm and 2 μm, respectively possesses a Q factor of $2.5 \times 10^7$ and filled with GFP solution of 20 μM concentration. We seal both ends of the microbubble by the UV glue so that the GFP solution does not volatilize. When not used, the microbubble is stored in a refrigerator at 3°C–4°C. We can see from Figs. 5 (a) and (b) that the threshold behavior and the optical spectrum of the GFP laser are rather stable after 23-day preservation; the threshold on the 1st and 23rd day are 1.35 μJ/mm$^2$ and 1.24 μJ/mm$^2$, respectively. The lasing spectra for these two days, shown in Fig. 5 (c), (d) also bear similar result. Notably, this durability may even last longer, which we have not tested. Due to the low threshold characteristics of our laser, the irradiation damage of the pump light to the GFP, the gain medium of the laser is minimal to ensure the GFP solution has a long-term stability. Such high durability is of great significance in practical applications, which has not been proved thus far.

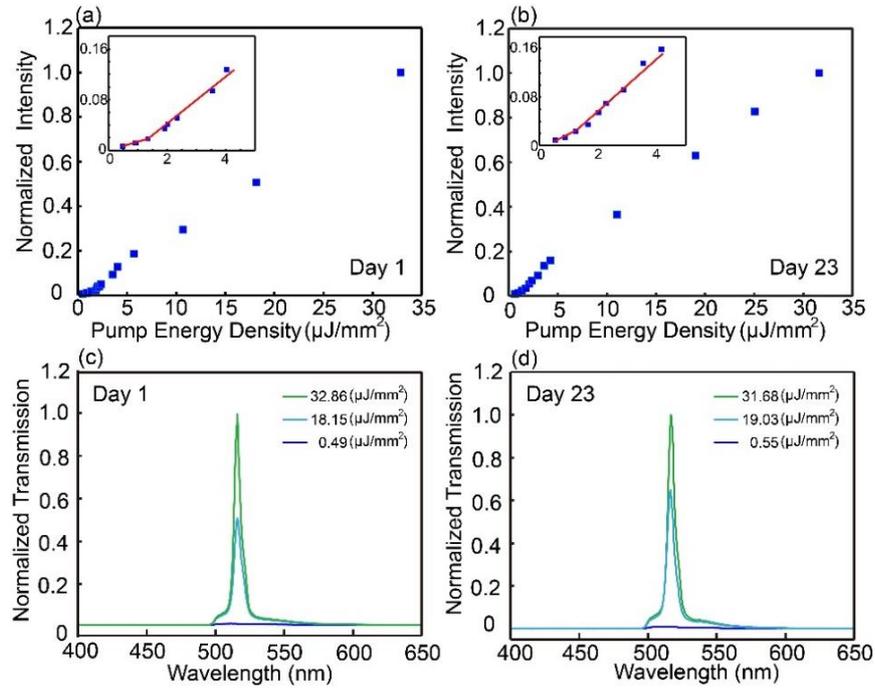

Fig. 5. (a) Normalized laser intensity with the variation in the pump energy density on the first day, inset: close up of the low pump energy regime showing the threshold of 1.35 μJ/mm$^2$; (b) Normalized laser intensity with the variation in the pump energy density 23 days later, inset: close up of the low pump energy regime with the threshold of 1.24 μJ/mm$^2$; (c) Lasing spectra on the first day under different pump energy densities; (d) Lasing spectra 23 days later under different pump energy densities

From the lasing threshold behavior, we further develop a GFP concentration detector in the microbubble resonator. We utilize a microbubble with the outer diameter and wall thickness of 90 μm and 2 μm, whose Q factor is $1.3 \times 10^7$ when filled with GFP solution. We test the lasing threshold at different GFP concentrations. Fig. 6 (a), (b), and (c) plots the lasing threshold at the GFP concentrations of 10 μM, 9.09 μM, and 8.70 μM respectively, while the lasing spectra is illustrated in Fig. 6 (d), (e), and (f). The corresponding lasing thresholds are 2.25 μJ/mm$^2$, 3.26 μJ/mm$^2$, 3.63 μJ/mm$^2$. There is a decrease in the threshold energy with an increase in the GFP concentration. As such, the microbubble based GFP laser is a sensitive detector for determining GFP concentrations. According to the relationship between the laser threshold and the protein concentration, we can distinguish two GFP solutions with concentration difference higher than 0.39 μmol/L for the high-precision measurement of the solution concentration. This concentration detection test preliminarily confirmed the great application potential of our laser system in the field of biological detection.

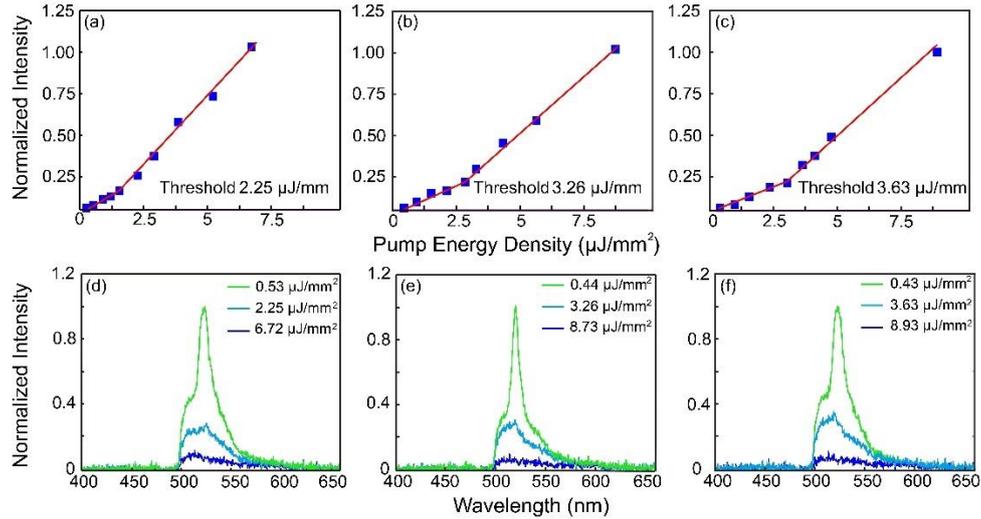

Fig. 6. (a) Normalized laser intensity with the variation in pump energy densities, the concentration of GFP is 10 μM; (b) Laser intensity with the variation in pump energy densities, the concentration of GFP is 9.09 μM; (c) Laser intensity with the variation in pump energy densities, the concentration of GFP is 8.70 μM; (d) Lasing spectra under different pump energy densities, GFP concentration 10 μM; (e) Lasing spectra under different pump energy densities, GFP concentration 9.09 μM; (f) Lasing spectra under different pump energy densities, GFP concentration 8.70 μM

## 4. Conclusion

In conclusion, we have successfully developed a miniature ultralow threshold GFP laser by using the high-quality factor microbubble resonator with the lowest energy threshold and the highest quality factor among all known fluorescent protein lasers of the similar type. The ultralow pump threshold of this laser can effectively avoid the damage of biological gain mediums by light to maintain good luminescence performance for a long time. To verify this, we tested the long-term stability of the GFP laser. The results show that the fluorescent light amplification effect can last for at least 23 days at 3°C–4°C, which lays a solid foundation for the practical application of the fluorescent protein laser. Finally, we found that there is a high dependence between the laser threshold and protein concentration, which indicates that this biological laser can be used for ultrasensitive detection of fluorescent protein concentration. This work shows the potential of microcavity biological laser in biological detection technology. Previously, our team has carried out a series of research work in protein structure calculation [40-42]. On this basis, we can further design proteins with high fluorescence quantum yield in different wavelengths as the gain medium of optical microbubble laser, so as to prepare for the final development of high-performance protein laser, and use them for temperature and pH sensing, ultrasensitive detection of biomolecules, and other technologies. Besides, we will explore the possibility to use other type of resonators such as microtoroid [43, 44] and microdisk [45] so as to meet different context of use.

**Funding.** China National Postdoctral Program for Innovative Talentents (BX20200057); National Natural Science Foundation of China (61771278);

**Acknowledgment.** Z. Yin acknowledges financial support from Beijing Institute of Technology Research Fund Program for Young Scholars

**Disclosures.** The authors declare no conflicts of interest.